%% file: Identifiers.tex
\documentclass{ivoa}
\input tthdefs

\SVN$Rev: 3158 $
\SVN$Date: 2015-11-20 09:04:09 +0100 (Fri, 20 Nov 2015) $
\SVN$URL: https://volute.g-vo.org/svn/trunk/projects/registry/Identifiers/Identifiers.tex $

\newcommand{\abnfterm}[1]{%
  \ensuremath{\,\hbox{\texttt{\char'042\relax #1\char'042}}\,}}
\newcommand{\abnfrepeat}[1]{\,*#1}

\newcommand{\abnfor}{\ensuremath{\,\,/\,\,}}
\newcommand{\abnfnt}[1]{\ensuremath{\langle\textit{#1\/}\rangle\,}}
\newcommand{\abnfto}{=}

\iftth
\renewcommand{\abnfterm}[1]{%
  #1}
\renewcommand{\abnfnt}[1]{#1}

\fi

\title{IVOA Identifiers}

\ivoagroup{Resource Registry}

\author[http://www.ivoa.net/twiki/bin/view/IVOA/MarkusDemleitner]{Markus Demleitner}
\author[http://www.ivoa.net/twiki/bin/view/IVOA/RayPlante]{Raymond Plante}
\author[http://www.ivoa.net/twiki/bin/view/IVOA/TonyLinde]{Tony Linde}
\author[http://www.ivoa.net/twiki/bin/view/IVOA/RoyWilliams]{Roy Williams}
\author[http://www.ivoa.net/twiki/bin/view/IVOA/KeithNoddle]{Keith Noddle}
\author{and the IVOA Registry Working Group}

\editor{Markus Demleitner}

\previousversion[http://www.ivoa.net/Documents/REC/Identifiers/Identifiers-20070302.html]{REC-1.12}
\previousversion[http://www.ivoa.net/Documents/PR/Identifiers/Identifiers-20050302.html]{PR-20050302}
\previousversion[http://www.ivoa.net/Documents/PR/Identifiers/Identifiers-20040621.html]{PR-20040621}
\previousversion[http://www.ivoa.net/Documents/WD/Identifiers/Identifiers-20040209.html]{WD-20040209.html}
\previousversion[http://www.ivoa.net/Documents/PR/Identifiers/Identifiers-20031031.html]{WD-20031031}
\previousversion[http://www.ivoa.net/Documents/WD/Identifiers/Identifiers-20030930.html]{WD-20030930}
\previousversion[http://www.ivoa.net/Documents/WD/Identifiers/Identifiers-20030830.html]{PR-20030830.html}
       
\begin{document}
\begin{abstract}
An IVOA Identifier is a globally unique name for a resource
within the Virtual Observatory.  This
name can be used to retrieve a unique description of the resource
from an IVOA-compliant registry or to identify an entity like a dataset
or a protocol without dereferencing the identifier.  
This document describes the syntax
for IVOA Identifiers as well as how they are created.  
The syntax has been defined to encourage global-uniqueness naturally
and to maximize the freedom of resource providers to control the
character content of an identifier.
\end{abstract}

\section*{Acknowledgments}

This document builds on the concept of a Uniform Resource Identifier
as described in RFC 3986 \citep{std:RFC3986} and its predecessors.

This document has been developed with support from the
\href{http://www.nsf.gov}{National Science Foundation}'s
Information Technology Research Program under Cooperative Agreement
AST0122449 with The Johns Hopkins University, from the
\href{http://www.pparc.ac.uk}{UK Particle Physics and Astronomy
Research Council (PPARC)}, from the
\href{http://fp6.cordis.lu/fp6/home.cfm}{European Commission's Sixth
Framework Program} via the \href{http://www.astro-opticon.org/} {Optical
Infrared Coordination Network (OPTICON)}, and from the German
Astrophyiscal Virtual Observatory GAVO, BMBF grant 05A14VHA.

\section*{Conformance-related definitions}

The words ``MUST'', ``SHALL'', ``SHOULD'', ``MAY'', ``RECOMMENDED'', and
``OPTIONAL'' (in upper or lower case) used in this document are to be
interpreted as described in RFC 2119 \citep{std:RFC2119}.

\section*{Usage of ABNF}

This specification uses ABNF \citep{std:RFC2234} to specify grammar
rules.  The rules from RFC 3986 are assumed throughout.  Where both this
specification and RFC 3986 define a nonterminal, the rule in this
specification overrides the corresponding rule from RFC 3986.

For explicitness, we write ABNF nonterminals in angle brackets
(\abnfnt{like this}) throughout.

\section{Introduction}

Virtual Observatory applications frequently need to
unambiguously refer to some resource or concept
which is described elsewhere.  It is therefore necessary to
define global, potentially
dereferenceable identifiers.  In the VO, these are called
IVOA identifiers (IVOIDs).
An unambiguous reference within the entire Virtual Observatory
requires that the identifier is globally unique.  Ensuring
this uniqueness inevitably requires oversight by a moderating
authority; however, a flexible framework can minimize the opportunity
for duplicated identifiers.

Many data providers in the VO were creating
and using identifiers long before this specification was developed.  
Their choices of identifiers were made
presumably to best fit the needs of the data.  
In order to minimize the cost of adoption of the IVOA identifier framework
the design specified here maximizes the
control providers have, thus allowing the reuse of the identifiers 
data providers already have in place
as well as the creation of 
new idenfiers that are consistent with their overall
organization.  

Identifiers are crucial to the operation of registries that aid users in
discovering data and services \citep{std:RI1}.  In general, a registry stores
descriptions of data and services in a searchable form, and it
distinguishes them by the unique identifier defined here.  It thus
serves as a primary key for the VO Registry, and thus allows
dereferencing identifiers to metadata about a resource (the
resource record).

IVOA identifiers with query or fragment parts can furthermore
reference essentially arbitrary
entities like datasets or protocols, based on this primary mechanism of
dereferencing.

We recognize that resources do
not always remain in the control of a single organization forever.
This
necessitates a form of referencing that is
location-independent -- or more precisely, organization-independent.
Apart from enabling seamless transfers of data curation, an
attractive use case for such identifiers is 
when several copies of a dataset exist at several locations around
the VO and one could refer to all of them collectively, deferring the choice
of a particular instance until it is actually needed.
Such references thus serve as 
\emph{persistent} pointers to data that can be flexibly resolved.
This is very important to journal publishers
that wish to refer to data in publications (whose useful life might be
measured in decades) without worry that the references will become
obsolete.

This specification, in contrast, defines 
\emph{organization-dependent identifiers}.
Persistent, organization- and location-independent identifiers are
\emph{not} (directly) defined here.

Referencing resources is
addressed by the IETF standard for URIs, RFC 3986 \citep{std:RFC3986}.
Thus, the framework proposed
in this document builds directly on this standard.  Essentially, this
standard sets the parameters left open for application use
by RFC 3986.

\subsection{Definitions}

A \emph{Uniform Resource Identifier}
(URI) is defined by RFC 3986 as ``a compact sequence of
characters that identifies an abstract or physical resource'' which
complies with the syntax specification of that document
\citep{std:RFC3986}.  It can point to an
actual retrievable resource, but there is no requirement for it to be
dereferenceable at all, let alone by a stock web browser.

An \emph{IVOA identifier}, or IVOID, is a
special sort of URI complying with all parts of this specification.
Historically, these have also been known as \emph{IVOA Resource Names}
(IVORNs), in parallel to
the \emph{Uniform
Resource Names} (URNs) that formulated extra requirements on persistency
and location-independence.  As plain IVOIDs do not fulfill those 
requirements and the term URN has been deprecated by
RFC 3986, we now deprecate the term IVORN, too.

A full IVOID can thus be split into a \emph{Registry part} (schema,
authority, and path) and a possibly empty
\emph{local part} consisting of query and fragment component, again
using RFC 3986 nomenclature.  An IVOID with an empty local part is also
known as a \emph{Registry reference}.

In VO practice, the term \emph{resource} is somewhat ambiguous.
The IVOA Recommendation on
Resource Metadata \citep{std:RM}, from here on referred to as RM,
defines it as ``a VO element that can be described in terms of who
curates or maintains it and which can be given a name and a unique
identifier.''  It then goes on to define the relevant pieces of
metadata, which later provided the foundations of the data model behind
the IVOA Registry.

This might lead to the expectation that there is a 1:1 relationship
between Registry records, ``VO resources'', and IVOA identifiers, and
version 1 of this document essentially implied as much.  In this
version, we only require Registry references to resolve in the
Registry.

IVOIDs having a nonempty local part do not dereference to
Registry records.  Since we want to
maintain the notion that a resource is whatever a URI points to,
``resource'' as used here does \emph{not} correspond to the usage of the term in
VOResource \citep{std:VOR}.  To maintain the distinction, we call 
resources in the sense of VOResource \emph{Registry records}. 
These form a subset of the resources (in the URI sense) that
can be referenced by IVOIDs.

We refer to organizations and providers in the sense that they
are defined in RM:

\begin{quotation}
An \emph{organization} is a specific
type of resource that brings people together to pursue
participation in VO applications.  Organizations can be hierarchical
and range greatly in size and scope.  At a high level, it could be a
university, observatory, or government agency.  At a finer level, it
could be a specific scientific project, space mission, or individual
researcher.  A \emph{provider} is an
\emph{organization} that makes data and/or services
available to users over the network.  
\end{quotation}

Definitions of other types of resources, including data collection
and service, are also provided in RM, and
are assumed by this document.

\subsection{Selected Requirements}

This proposal is the result of various requirement studies for VO
identifiers and registries in general (e.g.
NVO ID
requirements\footnote{
\url{http://web.archive.org/web/20070226120639/http://nvo.ncsa.uiuc.edu/~rplante/VO/metadata/oidreq2.txt}}).  
This section highlights a few of the important
ones that guided the design of the ID framework.

\begin{enumerate}
  \item A single framework should be used to identify anything a VO
       application can refer to, including organizations, projects
       (mission/telescope), data collections, and services.

  \item It should be easy to compare two instances of an identifier to
       determine if they refer to the same object.

  \item It should be possible to use an identifier to access a unique
       description of the resource it identifies.

  \item The framework should maximize the freedom of data providers to  
       choose identifiers for resources and collections under their  
       control.
\end{enumerate}

\subsection{Rationale for Version 2}

A need for revising the IVOA Identifiers specification was discerned
ever since \citet{note:uriforms} pointed out that common practices
regarding dataset identifiers were not in line with URI semantics.
Also, with the publication of StandardsRegExt
\citep{std:STDREGEXT}, it became advisable to regulate the ways
standards are referenced in the VO in ways compatible with the spirit of
that standard.

As the Registry Working Group set about revising the Identifiers
recommendation, it was decided to drop the XML representation for
IVOIDs since it complicated the text but had never actually been found
useful.  Even in XML serializations only the URI form of IVOA
identifiers had been used.  Dropping the XML form nevertheless
constitutes an incompatible change, which necessitates an increase in the
major version number.

Despite the new major version, consensus was that current usage of
IVOIDs should not be impacted and existing practices sanctioned as far
as possible.  Apart from deprecating the use of fragment identifiers to
distinguish datasets and restricting authorities to only use
\abnfnt{unreserved} characters (which does not impact existing authority
identifiers), this specification therefore refrains from modifying
version 1 regulations even where they were found somewhat burdensome
(e.g., as regards case-insensitiveness in resource keys).

The opportunity of a revision was also used to organize the
specification content in parallel to RFC 3986; for instance, the notion
of stop characters from version 1 -- necessitated by the non-URI XML
representation -- has no counterpart in non-IVOID URIs and is now
encompassed naturally by the usual rules for parsing URIs.

Closely following RFC 3986 also allows rigorous definitions for the
interpretation of local parts.  In addition to what version 1 
specified, we now allow
percent-encoded characters there, and we comment on techniques to
resolve IVOIDs with such local parts.  This is finally used to define
the standard and the dataset identifiers that started the revision
process.

\subsection{IVOA Identifiers within the VO Architecture}

\begin{figure}[ht]
\centering
\includegraphics[width=0.9\textwidth]{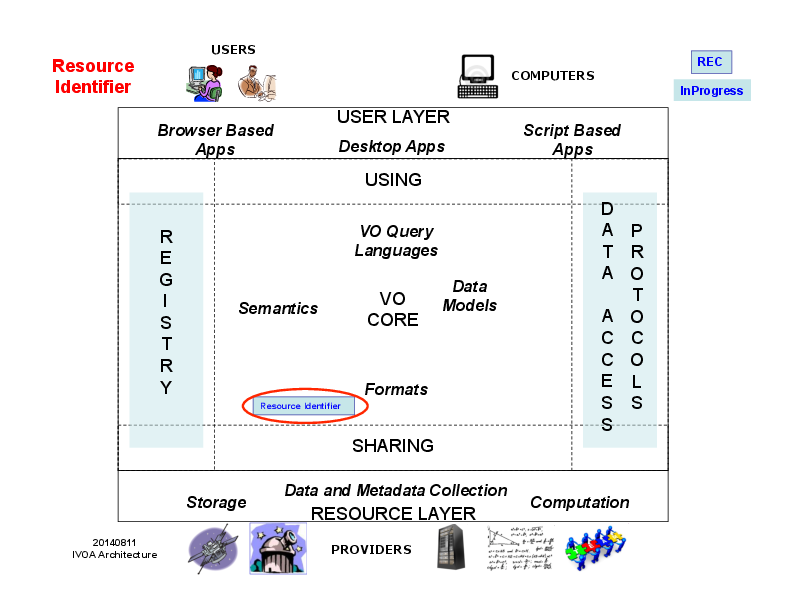}
\caption{Architecture diagram for the IVOA Resource Identifier
specification}
\label{fig:archdiag}
\end{figure}

Fig.~\ref{fig:archdiag} shows the role this document plays within the
IVOA architecture \citep{note:VOARCH}.  As identifiers are the primary
keys into the Registry, essentially all standards regulating the
Registry depend on this specification.  The data access protocols are
mainly impacted through the use of dataset identifiers -- e.g., in SSAP
\citep{std:SSAP} and Obscore \citep{std:OBSCORE} --, which are also
IVOID.  For the same reason, VOEvent is impacted.

The core of this standard has no dependencies on other VO standards.
The section on identifiers for standards depends on
StandardsRegExt \citep{std:STDREGEXT}.

\section{Specification}

After a brief, informal specification that should be enough for
non-demanding applications, this section gives, for each relevant part
of RFC 3986, additional requirements for IVOA identifiers.  The
normative content should be read together with RFC 3986.

\subsection{Overview (non-normative)}

IVOA identifiers (IVOIDs for short) are RFC 3986-compliant URIs with a
scheme of \texttt{ivo}.  Thus, their generic form is

$$\underbrace{\texttt{ivo:}
\texttt{//}
\abnfnt{authority}
\abnfnt{path}}_{\mbox{Registry part}}
\underbrace{\texttt{?}\abnfnt{query}
\texttt{\#}\abnfnt{fragment}}_{\mbox{local part}},
$$
where \abnfnt{path} is either empty or starts with a slash, and both
items in the local part are optional.
.

IVOIDs consisting of only scheme and authority are known as authority
identifiers and play a special role in creating other IVOIDs (see
sec.~\ref{sect:creating}).   IVOIDs without a local part 
must resolve to a Registry record within the IVOA Registry.
Likewise, for all IVOIDs, the IVOID resulting from stripping the local
part (the Registry part) must resolve within the IVOA Registry.  It is
called a \emph{Registry reference}.

The RFC 3986 \abnfnt{path} element is called \emph{resource
key} in IVOIDs.

Authority ids must consist of letters, numbers, dashes and dots
exclusively.   Resource keys must not contain URI reserved characters
(essentially, only alphanumeric characters, dashes, dots,
underscores, and tildes are allowed) except where an IVOA standard
defines how they are to be treated.

The Registry references are,
as a whole, compared case-insensitively, and must be treated
case-insensitively throughout to maintain backwards compatibility with
version 1 of this specification.  When comparing full IVOIDs, the local
part must be split off and compared preserving case, while the registry
part must be compared case-insensitively.

To make IVOIDs useful where these complex rules are hard to implement
(e.g., database columns), handling applications SHOULD NOT change the
case of any part of IVOIDs when these might have a local part.

Examples for IVOIDs:

\begin{itemize}
\item \nolinkurl{ivo://ivoa.net} -- an IVOID without a resource key,
i.e., an authority; dereferencing in the Registry must yield a
\xmlel{vr:Authority}-typed record.

\item \nolinkurl{ivo://ivoa.net/std/Identifiers} -- an IVOID with a
resource key.  Dereferencing this in the Registry must yield a resource
record. As long as there is no local part, an IVOID only differing 
in case, e.g., 
\nolinkurl{ivo://IVOA.NET/std/identifiers}, is in every respect equivalent to
it.

\item \nolinkurl{ivo://example.org/~?path/to/\%C3\%89CLAIRE} -- an IVOID
without guarantees as to if it resolves and what it resolves to.  The
Registry reference \nolinkurl{ivo://example.org/~} must resolve to a valid Registry
record, though.

\item \nolinkurl{ivo://example.org/svc?voc.xml#Term} -- an IVOID
conceptually referencing some item within
\nolinkurl{ivo://example.org/svc?voc.xml}.  If that latter IVOID can be
dereferenced, there should be an entity within the resource retrieved
that is itself identified by \texttt{Term}.  The classic example would
be an element with an \xmlel{id} of \texttt{Term} within an XML
document.
\end{itemize}

The remainder of this section contains a formalization of these points.

\subsection{Characters}

\label{sect:chars}

This specification poses no additional global constraints on the
character content of IVOIDs over what Section~2 of RFC 3986 specifies.
Special restrictions on the authority part and the resource key are
given below.  In particular, the \abnfnt{gen-delims} have, where
applicable, the standard URI interpretation.  As IVOIDs have no use for
IPv6 addresses or user components, square brackets and the commercial at
sign MUST NOT occur literally in IVOIDs anywhere.

The \abnfnt{sub-delims} MUST NOT be part of the resource key unless
another IVOA specification defines their use.  Their use in local parts
is not restricted by this specification, nor is any semantics defined
for them.  Other IVOA specifications may furnish them with semantics.

In IVOIDs, characters from \abnfnt{unreserved} MUST NOT be
percent-encoded.

Percent-encoded characters are allowed in local parts (but neither in
authority nor the resource key).  When
specifications or applications require text to be percent-encoded within
an IVOID, the text MUST be encoded in UTF-8.

\subsection{Syntax Components}

\subsubsection{Scheme}

The \abnfnt{scheme} part of IVOIDs is \texttt{ivo}.  Note that, by RFC
3986, scheme identifiers are case-insensitive.

A URI that uses this scheme (an IVOID) signals that:

\begin{itemize}

\item the registry part of the IVOID
  and the resource it refers to have been
  registered in the VO Registry
\item the URI complies with the additional restrictions laid down in
this document
\end{itemize}

The ivo scheme does not imply a transport protocol by which the resource
may be accessed.  Agents, in general, should not depend on implicit
mappings between IVOIDs and URIs in other schemes like \texttt{http}
when dereferencing them.  The only defined way to dereference IVOIDs is
described in sect.~\ref{sect:dereferencing}.  Resource publishers,
however, may support additional mappings between identifiers and other
URIs (such as http URLs) that they manage; in this case, agents should
only assume the mapping applies within the domain of the publisher.

\subsubsection{Authority}

\begin{admonition}{Note}
While the syntax for the authority identifiers
       allows it to look just like a DNS hostname, current convention
       discourages this practice to avoid the suggestion that an IVOA
       Identifier can be resolved like a common http URL.  
       As of this writing, the
       convention of the US Virtual Astronomical Observatory (VAO)
       is hierarchical naming that
       combines the publishing organization name with the project or
       archive (e.g. ``adil.ncsa'') while leaving out fields like
       ``.edu''
       or ``.org''.  In the AstroGrid
       project, the convention is to use a DNS name in reverse order
       (e.g. ``org.astrogrid.www''); this practice has the advantage of
       reducing the probability that two organizations will want to
       use the same authority identifier.
\end{admonition}

A \emph{naming authority} is an
organization (usually a data
provider) that has been granted the right by
the IVOA to create IVOA-compliant identifiers for resources it
registers.  See sect.~\ref{sect:creating} for
details on how this right is granted.  The naming authority creates
IVOIDs with empty local parts within the scope of one or more 
authority identifiers.

The \emph{authority} component of an IVOID is severely restricted over
RFC 3986 as follows:

\begin{itemize}
\item it MUST be at least three characters long
\item it MUST begin with an alpha-numeric character
\item it MUST NOT contain percent-encoded characters
\item it MUST NOT contain characters outside of \abnfnt{unreserved},
with the tilde strongly discouraged
\item there are no \abnfnt{userinfo} or \abnfnt{port} components
\end{itemize}

In ABNF, using the symbols from RFC 3986, an authority identifier
in IVOIDs thus has the form:

\begin{eqnarray*}
  \abnfnt{authority} &\abnfto& \abnfnt{alphanum} \abnfnt{unreserved}
    \abnfnt{unreserved} \abnfrepeat{\abnfnt{unreserved}}
\end{eqnarray*}

A naming authority is allowed to control multiple
authority identifiers to organize related resources into different
namespaces.  For example, an organization may
choose to control two authority identifiers, one for research-related
resources and one for education/outreach resources, even though they
are all maintained by the same organization and perhaps made available
through the same machine.

\paragraph{Examples for valid authorities}

\begin{compactenum}[(1)]
\item \texttt{nasa.heasarc}
\item \texttt{n\_1a.alph-0.02}
\item \texttt{123} (authorities can start with a number)
\end{compactenum}

\paragraph{Examples for invalid authorities}

\begin{compactenum}[(1)]
\item \texttt{a2} (less than three characters)
\item \texttt{\_temporary.id} (authorities must begin with an alphanumeric
character, which the underscore is not)
\item \texttt{DAT\%41} (percent-encoded characters are not allowed, even if they
work out to be unreserved characters)
\item \texttt{de!uni-hd!physics\#ari} (not entirely consisting of unreserved
characters)
\end{compactenum}

\subsubsection{Resource Key}

\label{sect:reskey}

RFC 3986's \abnfnt{path} part of an IVOID is called a \emph{resource key}.
It is a 
name for a resource that is unique within the namespace of an
authority identifier.  The naming authority creates keys for its namespaces
and has complete control of their forms beyond the syntax constraints
specified here.

On top of the definitions in RFC 3986 for paths, section 3.3, resource keys in
IVOIDs are further constrained in that

\begin{itemize}
\item \abnfnt{segment} MUST NOT contain percent-encoded characters 
\item \abnfnt{segment} MUST NOT contain colons or commercial at signs
\item Only \abnfnt{path-abempty} expansions are allowed
\end{itemize}

In ABNF, using or overriding the symbols of RFC 3986, this means:

\begin{eqnarray*}
  \abnfnt{path} &\abnfto &\abnfnt{path-abempty}\\
  \abnfnt{segment} &\abnfto & \abnfrepeat{\abnfnt{ivo-segment-char}}\\
  \abnfnt{ivo-segment-char}& \abnfto&  \abnfnt{unreserved} \abnfor
    \abnfnt{sub-delims}
\end{eqnarray*}

Naming authorities MUST NOT create path
segments matching either ``.'' or ``..''; empty
segments, resulting in two or more consecutive slashes or a trailing
slash, are also forbidden.  In particular, as
described in sect.~\ref{sect:comparing},
such segments would not have the
special meaning they have in traditional file system pathnames; that
is, a resource key cannot be transformed by removing any kinds of
segments and still reference the same resource.

Note that, as discussed in sect.~\ref{sect:chars}, characters from
\abnfnt{sub-delims} MUST NOT be used in resource keys unless their
semantics is defined in an IVOA specification.  As percent-encoded
characters are not allowed in resource keys, these characters MUST NOT
occur in generic Registry references at all.

The naming authority is free to create a
resource key that suggests something about the resource it refers to.
Any meaning that is suggested by the resource key is intended only for
human consumption.  The character content of a resource key is not
semantically machine-interpretable within the context of the IVOA as
defined by this document.

The presence of a resource key is optional.  An identifier that
contains only an authority identifier refers to the authority
itself and MUST resolve to a \xmlel{vr:Authority}-typed resource record
\citep{std:VOR} in the IVOA Registry.

VO applications MUST be case-insensitive when processing 
resource keys.  In presentation,
the preferred use of case is set by the rendering of the key by the
naming authority when the IVOID is registered.  This may contain
capital letters to improve readability.

\paragraph{Examples for valid resource keys}

\begin{compactenum}[(1)]
\item \texttt{""} (i.e., the empty string; zero repetitions of (\abnfterm/ \abnfnt{segment}) are
legal)
\item \texttt{/reskey}
\item \texttt{/\char127 user/STScI\_1/1a-7z.u} (unreserved characters are
allowed, and arbitrarily many segments are allowed)
\end{compactenum}

\paragraph{Examples of invalid resource keys}

\begin{compactenum}[(1)]
\item \texttt{/} (empty \abnfnt{segment}s are forbidden)
\item \texttt{reskey} (nonempty resource keys must always start with a
slash)
\item \texttt{/data/} (empty \abnfnt{segment}s are forbidden)
\item \texttt{/data//other} (empty \abnfnt{segment}s are forbidden)
\item \texttt{/data/c/../d} (\abnfnt{segment}s that indicate tree traversal in
other URI schemes are forbidden)
\item \texttt{/data!g-vo.org} (although this might become legal when some
IVOA standard gives the bang -- which is from \abnfnt{sub-delims} -- an
extra meaning)
\item \texttt{/user/M\%fcller} (percent encoding is forbidden in resource
keys; if it were, the codepoint 0xfc is not in
\abnfnt{ivo-segment-char}; if that were true, it would still not be
valid utf-8)
\end{compactenum}

\subsubsection{Query}
\label{sect:querypart}

This specification does not pose constraints on \abnfnt{query} beyond
the definitions in RFC 3986.  It also does not define any semantics.

Creators of IVOIDs are encouraged to adhere to URI semantics, i.e.,
IVOIDs with different query parts should refer to different resources.

To allow some resilience towards clients erronerously case folding the
query part, operators SHOULD NOT define IVOIDs referring to different
resources differing only by case in the query part.

Still, operators are not required to perform case folding on query
parts.  Therefore, applications MUST NOT change the case of characters
in query parts.

\paragraph{Examples for valid query parts}

\begin{compactenum}[(1)]
\item \texttt{par1=val1\&par2=val2} (the classic use for query parts in
HTTP URLs as, e.g., generated by browser forms)
\item \texttt{//..//!:??} (but sub-delims, slashes and question marks
are allowed here, as are strings looking like forbidden segments in
resource keys)
\item \texttt{\%C2\%B5\%20Her} (percent-encoding special characters is
legal, but outside of ASCII one has to use utf-8; this example works out
to be ``$mu$ Her'')
\item \texttt{\%3A\%5B\%5D} (while the generic delimiters \#, [,
and ] are not allowed in query parts literally, they can be included
in percent-encoded forms)
\end{compactenum}

\paragraph{Examples for invalid query parts}

\begin{compactenum}[(1)]
\item \texttt{:\#[] bad} (most generic delimiters are not allowed
literally in query parts, nor is the blank)
\item \texttt{\%B5\%20Her} (sequences of percent-encoded characters must
be valid utf-8 after decoding)
\end{compactenum}

\subsubsection{Fragment}

This specification does not pose constraints on \abnfnt{fragment}
beyond the definitions in RFC 3986.

Creators of IVOIDs are encouraged to adhere to URI semantics, i.e.,
fragment identifiers should be used to distinguish between different
entities within the same parent resource as discussed in
\citet{note:uriforms}.  The details of this process depend on the type
of document being retrieved.  See sects.~\ref{sect:dereferencing} and
\ref{sect:standards} for details.

Applications MUST NOT change the case of characters in fragments.

For examples for valid and invalid fragments, see the examples for query
parts in sect.~\ref{sect:querypart}

\subsection{Usage}

IVOIDs are used to identify resources in the general sense, i.e., they
might refer to datasets, abstract concepts, etc.;  their Registry 
parts, on the other
hand, MUST always be dereferenceable, i.e., resolve in the VO Registry.

No hierarchy is implied in any of the components.  Therefore, there are
no relative URIs for IVOA Identifiers.  In effect, this specification
overrides the rule in section~4.1 of RFC 3986 to become

$$
\abnfnt{URI-reference} \abnfto \abnfnt{URI}.
$$

\subsection{Reference Resolution}
\label{sect:dereferencing}

Registry references
can always be resolved to a Registry record by querying a
searchable registry, for instance, using RegTAP \citep{std:RegTAP}.
Clients will usually have some Registry endpoint URLs built in, more
are discoverable as described in \citet{std:RI1}.  In a full registry
with an OAI-PMH interface, the OAI-PMH \emph{GetRecord} operation
provides another means for obtaining the Registry record referenced by
an IVOID.

If an IVOID's Registry part does not resolve in the Registry, 
clients SHOULD assume it
is obsolete and that any IVOID built with it does not reference an
existing resource or entity either.

When dereferencing IVOIDs with query parts, applications should first
dereference the reference part to a registry record.  From that, a service
should be identified that can dereference the full IVOID.  Concrete
procedures may be given in IVOA specifications introducing certain
resource types.  One example for this is sect.~\ref{sect:dids}.

There is no mechanism that would allow applications to tell from
an IVOID's form whether or not it can be dereferenced in any special
way.  Any such information has to be obtained from the context the IVOID
is found in.

For resolving IVOIDs with fragment identifiers, applications would again
resolve the Registry part in the Registry.  In the presence of a query
component, it would be dereferenced as just discussed to obtain a basic
document, otherwise the basic document is the Registry record itself.
The entity referred to is then extracted from the basic document by
means specific to the document type; one example of such a prescription
is given in sect.~\ref{sect:standards}.

As there are no relative IVOIDs, most of RFC 3986's section~5 does not
apply here.

\subsection{Normalization and Comparison}
\label{sect:comparing}

An important use of identifiers is comparing two instances to
determine if they refer to the same resource.  This will most commonly
occur when using an identifier to look up the associated resource
description in a registry.

IVOID comparison is according to RFC 3986, section 6.2.2, with the
following additional regulations:

\begin{itemize}
\item As no hierarchy is implied in any IVOID part, no path segment 
normalization is ever performed on IVOIDs.
\item As IVOIDs must not percent-encode characters that do not need to
be encoded, no percent-encoding normalization is ever performed on
IVOIDs.
\item In addition to scheme and authority as in RFC 3986, in IVOIDs the 
resource key is also compared case-insensitively.  This means that 
Registry references can be case-folded for processing.
\end{itemize}

Note that neither query parts nor fragment identifiers may be compared
case-insensitively or normalized in any other way; allowing this would
severely impact their usefulness, as they, in general, refer to
case-sensitive entities like XML ids or file system paths.

No further normalizations are performed in IVOID comparison, i.e.,
sections 6.2.3 and 6.2.4 of RFC 3986 do not apply.

For instance, given the IVOID
$$\mbox{\nolinkurl{ivo://example.com/res/key1?par=U\%20Pic\#Part1},}$$ the
IVOID
$$\mbox{\nolinkurl{IVO://EXAMPLE.COM/RES/KEY1?par=U\%20Pic\#Part1}}
$$ must compare equal, while the following IVOIDs must compare
non-equal:

\begin{itemize}
\item \nolinkurl{ivo://example.com/res/key1?par=u\%20Pic\#part1}
(query part and fragment are non case-insensitive)
\item \nolinkurl{ivo://example.com/./res/key1?par=U\%20Pic\#Part1}
(no path normalization takes place, even if that were a legal IVOID)
\item \nolinkurl{ivo://example.com/res/key1?par=U\%20Pic} (fragment
identifiers may not be stripped off for comparison)
\item \nolinkurl{ivo://example.com/res/key1?par=U\%20Pic\&\#Part1}
(query parts are not parsed, and their interpretation as key/value pairs
is up to data providers)
\item \nolinkurl{ivo://example.com/res/\%6Bey1?par=U\%20Pic\#Part1}
(no normalization of percent encoding takes place)
\end{itemize}

In general, the string-based comparison of identifiers 
cannot determine definitively if two identifiers refer to different
resources.  While it is not intended that a Registry record is
registered multiple times with different identifiers, it is not
disallowed by this specification.  In particular, it is possible that
two resources with different identifiers may be mirrors of each other;
such a relationship can only be determined by examining the metadata
contained in the descriptions associated with each identifier.

This concludes the additional constraints and regulations for IVOIDs
over RFC 3986 compliant URIs.  The remainder of this document
standardizes certain aspects not in the scope of RFC 3986.

\section{Creating Identifiers}
\label{sect:creating}

An important aim of the process for creating identifiers is to ensure
uniqueness.  In the context of IVOA 
identifiers, ``unique'' means that a given identifier MUST NOT refer
to two different resources at any instant.  Furthermore, the
identifier SHOULD refer to at most one resource over all time; that
is, IVOIDs should not be reused for unrelated resouces.  Note that a
resource may potentially be dynamic (such as 'weather at telescope' or
'current version of the standard') -- here, there is a conceptually unique
resource, even though the content of it may change in time.

Another aim of the identifier creation process is to trace the
delegation of authority over the identifier.  
In practice, a Registry reference is created by 
an organization when registering a resource.
Thus, only recognized naming authorities (or
persons representing such organizations) may create Registry references.

The details of the service used to claim a 
naming authority is described in the IVOA Registry
Interfaces standard \citep{std:RI2}.

Once an organization is recognized as a naming authority, it is free
to register any number of resources with identifiers having an
authority identifier that they control.  No 
organization may create an identifier with an
authority identifier it does not control.  The naming
authority has full control over the creation of a 
resource key as long as it conforms to the syntax
and uniqueness constraints described in this specification.

Likewise, once a Registry reference is established, any number of IVOIDs may be
built using it (e.g., when publishing new datasets).  In this case, the
VO Registry is not involved, IVOID creation happens under the exclusive
control of the owner of the service or data collection the Registry
reference refers to.

\section{Special Identifier Types}
\label{sect:specials}

This section discusses some special classes of IVOIDs that reference
something other than Registry records and for which identifier forms for
one reason or other must or should be uniform across the different other
standards that define the resources referenced.

\subsection{Dataset Identifiers}
\label{sect:dids}

DAL standards like Obscore \citep{std:OBSCORE}, SSAP
\citep{std:SSAP}, or Datalink \citep{std:Datalink} need to reference
datasets. The SSAP standard defines these as ``an individual data object
usually including associated metadata.'' In astronomy, single images or
spectra are datasets, but tables or more complex data products might, at
the publisher's discretion, also be referenced as a single dataset.

A reference to a dataset is called a dataset identifier (DID), more
specifically publisher DID if the DID was assigned by the dataset's
publisher, and creator DID if the DID was assigned by the dataset's
author.  Various standards mandate that DIDs must be IVOIDs.  

Historically, DIDs were customarily formed by adding fragment
identifiers to Registry reference, a practice recommended in
SSAP in versions up to 1.1.
This definition was criticized in
\citet{note:uriforms} as a potential interoperability issue.

Therefore, this specification deprecates the regulation from SSAP 1.1.
Instead, DIDs in the VO now MUST use the query part to distinguish
datasets within one VO resource.  In short, the separator between
Registry reference and local part now must be the question mark rather than the
octothorpe.  A welcome side effect is that the fragment identifier can
now be used to reference sub-entities within the datasets.

An example for a dataset id (that should actually resolve according to
the scheme laid out below) is $$
\mbox{\nolinkurl{ivo://org.gavo.dc/\~?flashheros/data/ca92/f0065.mt}.}$$

Existing DIDs in services implementing SSAP up to 1.1 and Obscore 1.0
are not affected by these requirements and may be used until the
respective services are updated to newer standards.

Note that by this specification publishers have no obligation to ensure
continued access to datasets identified with PubDIDs. They are \emph{not}
by themselves 
persistent identifiers with guarantees on resolvability.  Their main
function is to provide globally unique identifiers for use in, e.g.,
federating responses from different services.

Publishers are, however, encouraged to declare at least one capability
of a protocol dealing with 
PubDIDs\footnote{At the time of this writing, Datalink, Obscore, and
SSA are IVOA recommended protocols allowing queries involving PubDIDs.
SIA \citep{std:SIAP} will, according to current
proposed recommendations, have an analogous facility in version
2.0.} in the resource record referenced by the Registry part of
a PubDID (i.e., the URI in front of the first question mark).  In that
way, clients can attempt to retrieve data based on
stand-alone PubDIDs by querying the
Registry for the ``embedding'' resource and seeing if it supports any
protocol they implement.

The definition of a proper resolver or resolution strategy is beyond the
scope of this standard.  Although services prototyping such funtionality have
been written\footnote{e.g., GAVO's global PubDID resolver at 
\url{http://dc.g-vo.org/glopidir}.}, we
maintain additional efforts are required outside of Registry to build a
reliable infrastructure on top of PubDIDs.

\subsection{Standard Identifiers}
\label{sect:standards}

In many VO standards, it is important to express adherence to a
set of constraints.  
Common examples include the declaration of the protocol --
and the version of the protocol -- that an endpoint implements in
VOResource's \xmlel{capability} element or a data model represented
with a TAP service in TAPRegExt.  The resource record such identifiers
reference is defined by StandardsRegExt \citep{std:STDREGEXT}.  As such
records typically describe multiple versions of a standard, and a single
standard may contain definitions of multiple different capabilities that
need to be discerned, the simple Registry Reference of the standard record usually is
not enough.

Therefore, StandardsRegExt records should define one
\xmlel{key} element for each such referenceable
entity.  The \xmlel{name} child of this key, denoting both the kind of
capability and the major and minor version, is then what is referenced
by the identifier as defined by StandardsRegExt, such that the complete
element will typically have the form
$$
\abnfnt{standard-ref}\abnfterm{\#}\abnfnt{key-name}\abnfterm{-}\abnfnt{version}
$$

For instance, the standard exampleProto might define both a
data model \texttt{model} and a query capability \texttt{query}.  In
its version 1.0, there would be two standard keys \texttt{model-1.0} and
\texttt{query-1.0}.  In a \xmlel{capability} element in another
resource's Registry record, support of the query capability would then
be declared with the IVOID
\texttt{ivo://ivoa.net/std/exampleProto\#query-1.0}, whereas a TAP
service exposing the model would contain a \xmlel{dataModel} element
with an \xmlel{ivo-id} attribute of
\texttt{ivo://ivoa.net/std/exampleProto\#model-1.0}.

As the exampleProto develops, new standard keys like
\texttt{query-1.1} or \texttt{query-2.0} are added.  Note that while ideally,
the version tags in the keys will correspond to the version of the
document that defines them, this is not a requirement.  Indeed, if the
underlying model has no incompatible changes, even exampleProto 2.0
might specify that its data model would remain 
\texttt{ivo://ivoa.net/std/exampleProto\#model-1.0}.  This allows clients
to easily discover all services they can operate.

Registry interfaces will typically offer some pattern matching
capability for comparing such identifiers.
Clients should use that feature
to ignore minor versions if appropriate -- by the IVOA's versioning
rules \citep{std:docSTD}, 
a generic client for version 1 of a protocol should be able to
operate all version 1 services, regardless of their minor versions, and
clients implementing multiple versions of a standard can entirely ignore
the version tag.  For instance, with RegTAP \citep{std:RegTAP}
an exampleProto 1.0 client would look for capabilities for which
$$
\texttt{standard\_id LIKE 'ivo://ivoa.net/std/exampleProto\#query-1.\%'}
$$
holds, whereas a client that speaks both versions 1 and 2 of the
protocol would look for capabilities with
$$
\texttt{standard\_id LIKE 'ivo://ivoa.net/std/exampleProto\#query-\%'}.
$$

\appendix

\section{Changes from Previous Versions}

\subsection{Changes from PR-2015-07-09}

\begin{itemize}
\item Now deprecating the term IVORN, as historical usage has been too
inconsistent.  Instead, there is now the ``Registry part'' of an IVOID,
and an IVOID that only has a registry part is called a Registry
reference.
\item More examples
\item No longer suggesting a concrete algorithm for PubDID resolution;
instead, clear encouragement to PubDID minters to point to appropriate
services from the Registry part of a PubDID.
\item Editorial changes
\end{itemize}

\subsection{Changes from 1.12}

\begin{itemize}
\item Removed the (unused) XML representation of Identifiers.
\item Rewrote the section on URI forms to more closely correspond to
the organization of RFC 3986.
\item Case-insensitive handling of IVORNs is now a MUST.
\item Now allowing percent-encoded items outside of the authority and
resource key.
\item Added rules for forming URI-compliant dataset identifiers
\item Added rules for forming StandardsRegExt-compliant standard
identifiers.
\item Empty path segments, as well as those consisting exclusively of
dots, are now forbidden rather than just discouraged.
\item Dropped the recommendation to present authority identifiers in
lower case.
\item Generally moved to IVOID as the abbreviation for IVOA identifier,
defined IVORN to be the part of an IVOID without a local part.
\item Removed some obsolete introductory material that has been
superseded by other standards.
\item Migrated to ivoatex source
\end{itemize}

\subsection{Changes from v1.10}

\begin{itemize}
\item Moved ``!'' from the discouraged list of
       characters to the reserved list, 
       thereby disallowing its inclusion in IVOA identifiers.
\item Clarified the list of characters disallowed in an authority ID by:
   \begin{itemize}
      \item explicitly disallowing URI-escaped sequences.
      \item listing as reserved characters only those characters
              that are allowed by the URI spec but disallowed by this
              one.
      \item Listed in a tip box the characters that are disallowed
              by the URI spec.
   \end{itemize}
       As before, the definition of the resource key
       refers to the same list of
       reserved characters as those disallowed.
  \item Fixed numerous links and references.
\end{itemize}

\subsection{Changes from v1.0}

\begin{itemize}
\item The prohibition of using ``+'' and ``='' within
 Identifier components has been dropped.
\item Recommendations for authority ID strings
       have been updated to match current practice in AstroGrid and the
       NVO.
\item In the example schema in App. A, the namespace was altered to conform
       with IVOA conventions.  A correction was also made to the
       allowed pattern for AuthorityIDType to properly comply with the XML
       specification defined in section 3.2.1.
\item  various clarifications based on reviewer comments
\end{itemize}

\subsection{Changes from v0.1}

\begin{itemize}
\item Resource key is now required except when referring to a naming
       authority itself.
\item support for DNS-like authority IDs clarified.
\item added role of \# and ? as ``stop'' characters in URI form.
\item dropped non-binding Appendix B: Recommended Mechanism for
       becoming a Naming authority.
\end{itemize}

\bibliography{ivoatex/ivoabib}

\end{document}

%% file: tthdefs.tex
\newif\iftth

%% file: ivoatexmeta.tex
\newcommand{\ivoaDocversion}{2.0}
\newcommand{\ivoaDocdate}{2016-05-23}
\newcommand{\ivoaDocdatecode}{20160523}
\newcommand{\ivoaDoctype}{REC}
\newcommand{\ivoaDocname}{Identifiers}